\documentclass[journal]{IEEEtran}
\IEEEoverridecommandlockouts
\usepackage{graphicx}
\usepackage[normalem]{ulem}

\usepackage[swedish,english]{babel}

\usepackage{tikz,amsmath,amssymb,amsfonts,bm}
\usepackage{pgfplots}
\newcommand{\U}{U}
\newcommand{\G}{{G}}
\newcommand{\Gp}{{G}_{\rm p}}
\newcommand{\Ep}{\vec{E}_{\rm p}}
\newcommand{\Hp}{\vec{H}_{\rm p}}
\newcommand{\Ap}{\vec{A}_{\rm p}}
\newcommand{\phip}{{\phi}_{\rm p}}

\newcommand{\Jt}{\vec{J}_{\rm tot}}
\newcommand{\vr}{\vec{r}}

\renewcommand{\vec}[1]{{\boldsymbol#1}}
\newcommand{\diff}{\mathrm{d}}
\providecommand*{\mrm}[1]{\mathrm{#1}}
\providecommand*{\vhat}[1]{\hat{\vec{#1}}}
\providecommand*{\iu}{\ensuremath{\mrm{j}}}
\providecommand*{\eu}{\ensuremath{\mrm{e}}}
\renewcommand{\Im}{\mathop{\mathrm{Im}}}
\renewcommand{\Re}{\mathop{\mathrm{Re}}}
\newcommand{\Om}{\Omega}
\newcommand{\ie}{\textit{i.e.}\/ }
\newcommand{\eg}{\textit{e.g.}\/ }

\newcommand{\R}{\mathbb R}

\newcommand{\Z}{\mathbb Z}

\newcommand{\We}{W_{\rm e}}
\newcommand{\Wm}{W_{\rm m}}
\newcommand{\Kzmn}{k_{{\rm z}mn}}

\newcommand{\Ktmn}{\vec{k}_{{\rm t}mn}}

\newcommand{\Ktnull}{\vec{k}_{{\rm t}00}}
\renewcommand{\Pr}{P_{\rm r}}
\newcommand{\unitz}{\overline{\overline{\mathbf{1}}}_{\rm z}}
\newcommand{\unitdyadic}{\overline{\overline{\mathbf{1}}}}

\DeclareMathAlphabet\mathbfcal{OMS}{cmsy}{b}{n}
\usepackage{tikz-3dplot}

\begin{document}

% paper title
\title{Q-factor and bandwidth of periodic antenna arrays over ground plane}

\author{Andrei Ludvig-Osipov
and
B. L. G. Jonsson
\thanks{
This work was supported by Vinnova in ChaseOn/iAA and Swedish Foundation for Stategic Research (SSF) in AM13-0011}
\thanks{The authors are with the School of Electrical Engineering and Computer Science, KTH Royal Institute of Technology, Stockholm SE-10044, Sweden (e-mail: osipov@kth.se; ljonsson@kth.se).}}

\maketitle

\begin{abstract}
In this paper, the Q-factor expression for periodic arrays over a ground plane is determined in terms of the electric current density within the array's unit cell.
The expression accounts for the exact shape of the array element.
The Q-factor formula includes integration only over a volume of an element in a unit cell, and can thus be efficiently implemented numerically.
The examples show good agreement between the proposed Q-factor, the full-wave-calculated tuned fractional bandwidth, and the input-impedance-based formula by Yaghjian and Best (2005) for the array of tilted dipoles and the loops array. 
\end{abstract}


\section{Introduction}
The Q-factor gives accurate prediction of the fractional bandwidth, when an antenna is not too wideband (\ie $Q\geq 5$)~\cite{Yaghjian+Best2005}.
Thus, expressing the Q-factor in terms of physical quantities of antenna provides an understanding of the connection between bandwidth and those quantities.
In particular, the Q-factor representation in terms of the electric current density~\cite{Vandenbosch2010,Gustafsson+Jonsson2015stored,Jonsson+Gustafsson2015} has been used to obtain Q-factor bounds for finite-sized antennas, subject to various constraints on \eg antenna shape, size, or radiation pattern~\cite{Gustafsson+etal2012,Vandenbosch2011,Cismasu+Gustafsson2014,Tayli+etal2018,Shi+etal2017,Gustafsson+Nordebo2013}.
As a step towards obtaining similar type of bounds for large array antennas, we have recently reported the Q-factor expression in terms of electric current density for unit cell representations of antenna arrays~\cite{Osipov+Jonsson2019}.
The primary goal of this paper is to extend the expression to a practically important case for arrays: the presence of a ground plane.

The unit-cell Q-factor of a strip dipole array was derived by Kwon and Pozar, where one propagating mode was assumed~\cite{Kwon+Pozar2014}.
We generalized in~\cite{Osipov+Jonsson2019} the free-space case to include arbitrary three-dimensional geometries of array elements, and to permit arbitrarily directed currents and multiple propagating modes.
Yaghjian and Best input-impedance-based formula~\cite{Yaghjian+Best2005}, shown to be applicable to array structures in~\cite{Kwon+Pozar2014,Osipov+Jonsson2019}, provides a simple way to estimate the Q-factor from the input impedance and its frequency derivative, however, it does not give a direct connection to the electric current density.
Such connection is instrumental in obtaining fundamental bounds.

This paper presents the derivation of the unit-cell Q-factor for arbitrarily-shaped PEC arrays over the ground plane.
We have used the method of images to express potentials and fields in order to evaluate stored energies.
The resulting Q-factor expression is a quadratic form in terms of a current density at an array element.
The numerical examples compare the Q-factor obtained by three different methods: the here proposed Q-factor, the input-impedance Q by Yaghjian and Best~\cite{Yaghjian+Best2005} and a Q-equivalent representing a full-wave-solver determined tuned bandwidth.

\section{Stored energies for ground-plane case}
\begin{figure}[tb]
  \centering
	\includegraphics[scale=0.8]{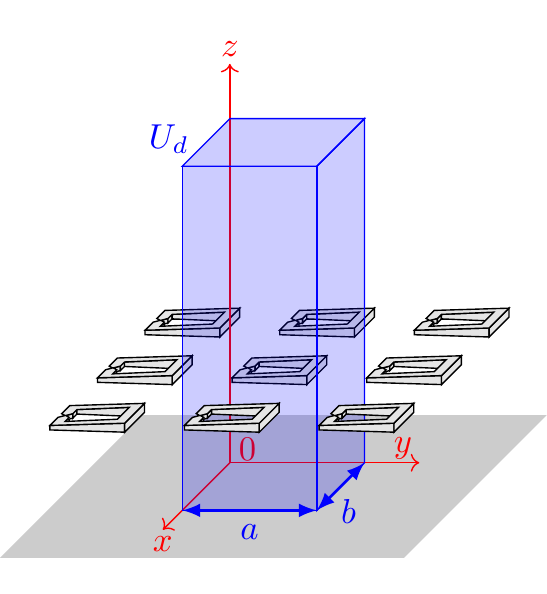}
  \caption{An example of an array over a ground plane with a unit-cell region represented by a blue column, and metal is shown by gray color.}
  \label{fig:array_sketch}
\end{figure}

Consider a phased array of PEC elements on a rectangular periodic grid over a ground plane.
The array elements are finite, in general, three-dimensional, arbitrarily shaped and regular enough to support a solution to Maxwell's equations.
An example of such a structure is shown in Fig.~\ref{fig:array_sketch}, where the metal regions are given by grey color.
The blue column represents a unit-cell region $U_{d}=\{(x,y,z)\in \R^3:x\in[0,a],y\in[0,b],z\in[0,d]\}$, where $a,b>0$ are the grid periods.
The phased array configuration with a phase-shift vector $\Ktnull$ is imposed by the condition on the current density
\begin{equation}
\vec{J}(\vec{r}+\vec{\zeta}_{mn})=\vec{J}(\vec{r})\eu^{\iu \Ktnull\cdot \vec{\zeta}_{mn}}, 
\label{eq:J_periodic}
\end{equation}
where $\vr\in\R^3$ is a position vector and $\vec{\zeta}_{mn}=am\hat{\vec{x}}+bn\hat{\vec{y}}$; $m,n\in \Z$.
The goal here is to find an expression for a Q-factor of such an array represented by the current density $\vec{J}$ in a unit cell.
This paper is entirely in frequency domain and the time-dependent phase factor $\eu^{\iu\omega t}$ is assumed but omitted.
The Q-factor is defined as~\cite{IEEEstandard2014}
\begin{equation}
    Q=\max(Q_{\rm e},Q_{\rm m}), \quad 
    Q_{\rm e/m} = \frac{2\omega W_{\rm e/m}}{P_{\rm d}},
\end{equation}
where $W_{\rm e/m}$ is electric/magnetic stored energy, and $P_{\rm d}$ is the dissipated power. For lossless radiating structures, the dissipated power is equal to the radiated power $P_{\rm r}$.

The ground-plane configuration is treated by the method of images.
The total current in the image problem is given by~\cite{Jin2011}
\begin{equation}
    \Jt(\vr)=\vec{J}(\vr) - \unitz\cdot\vec{J}(\vr_{\rm i}),
\end{equation}
where $\unitz = \vhat{x}\vhat{x} + \vhat{y}\vhat{y} - \vhat{z}\vhat{z}$ is a dyadic that inverts the sign of $z$-component of a vector, and the image coordinate is $\vr_{\rm i}=\unitz\cdot\vr$.
The vector and scalar potentials associated with the total current are (the Lorenz gauge is assumed~\cite{Jackson1999,Landau+etal1984})
\begin{equation}
\begin{split}
\vec{A}(\vec{r}_1)=
\mu\int_\Om  [\G(\vec{r}_1,\vec{r}_2) \unitdyadic  - \G(\vec{r}_1,\vec{r}_{\rm 2i})\unitz ] \cdot\vec{J}(\vr_2) \diff v_2,
\end{split}
\label{eq:A_G}
\end{equation}
\begin{equation}
\begin{split}
\phi(\vec{r}_1)=
\frac{-\iu}{\omega\epsilon}\int_\Om  (\nabla_2 [\G(\vec{r}_1,\vec{r}_2) - \G(\vec{r}_1,\vec{r}_{\rm 2i})])\cdot \vec{J}(\vr_2) \diff v_2.
\end{split}
\label{eq:phi_J}
\end{equation}
Here, $\G(\vec{r}_1,\vec{r}_2)$ is a free-space 2D-periodic Green's function~\eqref{eq:Greens_function}, $\mu$ and $\epsilon$ are the permeability and permittivity respectively. 
The electric and magnetic fields are found from the potentials via
\begin{equation}
\vec{E}=-\nabla \phi - \iu \omega \vec{A}, \quad
\vec{H} = \frac{1}{\mu} \nabla \times \vec{A}.
\label{eq:EH_phi_A}
\end{equation}
The electric and magnetic stored energies are given by subtracting the energy density of the propagating Floquet modes from the total energy density
\begin{equation}
\We = \frac{\epsilon}{4}
\int_{\U_\infty}|\vec{E}|^2-|\Ep|^2\diff v,
\label{eq:WeF}
\end{equation}
\begin{equation}
\Wm = \frac{\mu}{4}
\int_{\U_\infty}|\vec{H}|^2-|\Hp|^2\diff v.
\label{eq:WmF}
\end{equation}
Here $\Ep,\Hp$ are electric and magnetic fields associated with propagating Floquet modes.
They are obtained similarly to $\vec{E},\vec{H}$ with the Green's function $\G(\vec{r}_1,\vec{r}_2)$ replaced by its propagating-modes part $\Gp(\vec{r}_1,\vec{r}_2)$, see~\eqref{eq:Greens_function_F}.
Note that the energy densities are integrated above the ground plane.
A similar definition of the stored energies for the free space case is used in our earlier work~\cite{Osipov+Jonsson2019}.

We follow the same approach as in the free-space case~\cite{Osipov+Jonsson2019} to derive the stored energies.
The electric stored energy is represented by the potentials
\begin{equation}
\begin{split}
\We&=\frac{\epsilon}{4}\int_{\U_\infty} (|\nabla\phi|^2-k^2|\phi|^2) \\
&+ \omega^2(|\vec{A}|^2-|\Ap|^2) - k^2(|\phi|^2-|\phip|^2)\diff v.
\end{split}
\label{eq:WeF_potentials}
\end{equation}
Each pair in the integrand is then addressed separately to derive the current density representation.

The representation of the first pair in terms of the current density is found from the scalar potential~\eqref{eq:phi_J} and the continuity equation $\iu\omega\varrho=-\nabla\cdot\vec{J}$ as
\begin{equation}
\begin{split}
W_{\rm e,1} &= 
\frac{\epsilon}{4}\int_{\U_\infty}|\nabla\phi|^2-k^2|\phi|^2\diff v =
\frac{1}{4}\Re \{ \int_\Om \phi\varrho^*\diff v\} \\
&= 
\int_\Om \int_\Om  \vec{J}^*(\vec{r}_1) \cdot {\rm \bf K}_{\rm e,1}(\vec{r}_1,\vec{r}_2) \cdot \vec{J}(\vec{r}_2) \diff v_2 \diff v_1,
\end{split}
\label{eq:W1}
\end{equation}
where the dyadic kernel is
\begin{equation}
{\rm \bf K}_{\rm e,1}(\vec{r}_1,\vec{r}_2) = 
    \frac{\mu}{4k^2}\Re \left\{ \overline{\overline{\nabla}}_1\nabla_2 [\G(\vec{r}_1,\vec{r}_2) - \G(\vec{r}_1,\vec{r}_{\rm 2i})] \right\}.
\end{equation}
Here, $\overline{\overline{\nabla}}_1$ is a Jacobian with respect to $\vr_1$.

The vector-potential pair is evaluated by straightforward substitution of the magnetic vector potential~\eqref{eq:A_G} and the vector potential $\Ap$, associated with the propagating modes and found by replacing $\G$ with $\Gp$ in~\eqref{eq:A_G},
\begin{equation}
\begin{split}
W&_{\rm em,1} =
\frac{\epsilon\omega^2}{4} \int_{\U_\infty} (|\vec{A}(\vec{r})|^2-|\Ap(\vec{r})|^2)\diff v \\
&=
\int_\Om \int_\Om 
\vec{J}^*(\vec{r}_1)
\cdot {\rm \bf K}_{\rm em,1}(\vec{r}_1,\vec{r}_2) \cdot
\vec{J}(\vec{r}_2)
\diff v_1 \diff v_2, 
\end{split}
\label{eq:We2}
\end{equation}
with
\begin{equation}
    {\rm \bf K}_{\rm em,1}(\vec{r}_1,\vec{r}_2) =
    \frac{\mu k^2}{4}\left( g(\vec{r}_1,\vec{r}_2) \unitdyadic
    -  g(\vec{r}_{\rm 1i},\vec{r}_2) \unitz \right).
\end{equation}
Here, the function $g$ is defined by
\begin{equation}
\begin{split}
g&(\vec{r}_1,\vec{r}_2) =
\int_{\U_\infty}
\G^*(\vr,\vr_1)\G(\vr,\vr_2) - \Gp^*(\vr,\vr_1)\Gp(\vr,\vr_2) \\ 
&+ \G^*(\vr,\vr_{\rm 1i})\G(\vr,\vr_{\rm 2i}) - \Gp^*(\vr,\vr_{\rm 1i})\Gp(\vr,\vr_{\rm 2i})
\diff v.
\end{split}
\end{equation}
Note here that the integration is performed for the unit-cell column at $z\geq 0$.
Utilizing the $z$-symmetry of the Green's function, we can rewrite the integral as
\begin{equation}
\begin{split}
g&(\vec{r}_1,\vec{r}_2) = \\
&\int_{\U_\infty\cup \U_\infty^-}
\G^*(\vr,\vr_1)\G(\vr,\vr_2) - \Gp^*(\vr,\vr_1)\Gp(\vr,\vr_2) 
\diff v,
\label{eq:g_kernel}
\end{split}
\end{equation}
where $\U_\infty^-$ denotes the unit-cell column at $z\leq 0$.
We recognize the calculation of $g$ from~\cite{Osipov+Jonsson2019} to find the analytic expression~\eqref{eq:little_g_exact}.
The last term for the scalar-potential pair is obtained analogously:
\begin{equation}
\begin{split}
W&_{\rm em,2} =
\frac{\epsilon k^2}{4} \int_\U (|\phi|^2-|\phip|^2)\diff v \\
&=\int_\Om \int_\Om 
\vec{J}^*(\vec{r}_1)
\cdot {\rm \bf K}_{\rm em,2}(\vec{r}_1,\vec{r}_2) \cdot
\vec{J}(\vec{r}_2)
\diff v_1 \diff v_2, 
\end{split}
\label{eq:We3}
\end{equation}
with
\begin{equation}
    {\rm \bf K}_{\rm em,2}(\vec{r}_1,\vec{r}_2) =
    \frac{\mu }{4}\overline{\overline{\nabla}}_1\nabla_2 \left( g(\vec{r}_1,\vec{r}_2) 
    -  g(\vec{r}_{\rm 1i},\vec{r}_2)  \right).
\end{equation}
%\LJ{((I would probably extract $\mu$ out of K, but it is optional.))}
For lossless materials Poyntings theorem~\cite{Jackson1999} gives
\begin{equation}
\Wm = \We + \frac{1}{2\omega} \Im P_{\rm c},
\label{eq:Wm_Pc}
\end{equation}
where
\begin{equation}
P_{\rm c} = 
-\frac{1}{2}\int_\Om\vec{E}\cdot\vec{J}^*\diff v
\label{eq:Pc_EJ}
\end{equation}
is the complex power.
Substitution of electric field~\eqref{eq:EH_phi_A} and potentials~\eqref{eq:A_G} and~\eqref{eq:phi_J} into~\eqref{eq:Pc_EJ} gives
\begin{equation}
\begin{split}
P_{\rm c}
 =\int_\Om \int_\Om 
\vec{J}^*(\vec{r}_1)
\cdot {\rm \bf K}_{\rm P}(\vec{r}_1,\vec{r}_2) \cdot
\vec{J}(\vec{r}_2)
\diff v_1 \diff v_2, 
\end{split}
\label{eq:Pc_J}
\end{equation}
where
\begin{equation}
\begin{split}
    {\rm \bf K}_{\rm P}(\vec{r}_1,\vec{r}_2) &= \frac{\iu \eta}{2k} \overline{\overline{\nabla}}_1\nabla_2 [\G(\vec{r}_1,\vec{r}_2) - \G(\vec{r}_1,\vec{r}_{\rm 2i})] \\ &+
    \frac{\iu k \eta}{2} [\G(\vec{r}_1,\vec{r}_2) \unitdyadic  - \G(\vec{r}_1,\vec{r}_{\rm 2i})\unitz ].
\end{split}
\label{eq:K_Pc}
\end{equation}
We introduce the magnetic contribution to stored energy as
\begin{equation}
\begin{split}
W_{\rm m,1} = 
\int_\Om \int_\Om  \vec{J}^*(\vec{r}_1) \cdot {\rm \bf K}_{\rm m,1}(\vec{r}_1,\vec{r}_2) \cdot \vec{J}(\vec{r}_2) \diff v_2 \diff v_1,
\end{split}
\label{eq:Wm1}
\end{equation}
with
\begin{equation}
{\rm \bf K}_{\rm m,1}(\vec{r}_1,\vec{r}_2) = 
    -\frac{\mu}{4}\Re \left\{ \G(\vec{r}_1,\vec{r}_2) \unitdyadic  - \G(\vec{r}_1,\vec{r}_{\rm 2i})\unitz  \right\}.
\end{equation}
Then, the imaginary part of~\eqref{eq:Pc_J} is identified as 
\begin{equation}
\Im P_{\rm c} = 2\omega(-W_{\rm e,1} + W_{\rm m,1}).
\end{equation}
The electric and magnetic stored energies thus are
\begin{equation}
	\We = W_{\rm e,1} + W_{\rm em,1} - W_{\rm em,2},
\end{equation}
\begin{equation}
	\Wm = W_{\rm m,1} + W_{\rm em,1} - W_{\rm em,2}.
\end{equation}
The radiated power is found from the Poyntings theorem~\cite{Jackson1999} as
\begin{equation}
\Pr = \Re P_{\rm c}.
\end{equation}

\section{Numerical examples}
In this section we compare the here derived Q-factor for arrays over a ground plane with  a tuned bandwidth, recalculated into an equivalent Q-factor $Q_{\rm B}$, as described in the next paragraph. We also compare our results with the input-impedance-based formula by Yaghjian and Best~\cite{Yaghjian+Best2005} $Q_{\rm Z}$. 
In all the examples, the here proposed Q-factors $Q_{\rm e}$ and $Q_{\rm m}$ are computed based on our in-house Method of Moments code with RWG basis functions on a triangular mesh.  

The tuned fractional bandwidth $B(\omega)$ at each angular frequency~$\omega$ is obtained as in~\cite[Section VI]{Yaghjian+Best2005}. 
The input impedance from a full-wave simulation is matched for the frequency $\omega$ by a series lossless reactive element (inductor or capacitor).
The fractional bandwidth is then estimated from the matched reflection coefficient $\Gamma$ for a given threshold $\Gamma_0$ (-10 dB in all numerical examples here).
The tuned fractional bandwidth is then recalculated into the equivalent Q-factor by
\begin{equation}
Q_{\rm B} = \frac{2\Gamma_0}{B\sqrt{1-\Gamma_0^2}}.
\end{equation}
In the examples here, to compute $Q_{\rm B}$, we used the input impedance from a unit-cell frequency-domain simulation in CST Microwave Studio.

The Q-factor by Yaghjian and Best~\cite{Yaghjian+Best2005}, based on input impedance $(R+\iu X)$,  is calculated by
\begin{equation}
Q_{\rm Z}(\omega) = \frac{\omega}{2R(\omega)}
\sqrt[]{ [ R'( \omega ) ]^2 + 
[ X'( \omega ) + |X(\omega)|/\omega]^2}.
\end{equation}
Here, $(.)'$ is a derivative with respect to $\omega$.

\subsection{Tilted dipole array}
\begin{figure}[tb]
  \centering
  \begingroup
	\tikzset{every picture/.style={scale=1}}
	\includegraphics{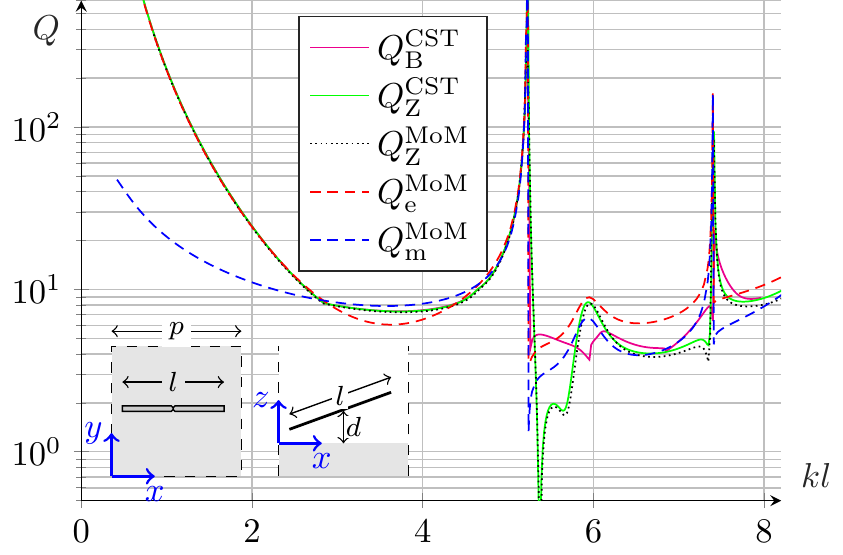}
  \endgroup
  ~
  \vspace{-12pt}
  \caption{The Q-factor of an array of tilted dipoles over a ground plane, dipole width $l/40$, $p=1.2l$, $d=0.25l$, angle $\alpha=20^\circ$ with ground plane.}
  \label{fig:Q_tilted_dipole}
\end{figure}

In the first example, we test the derived expression on an array of tilted dipoles.
The length of each dipole element is $l$, the width $w=l/40$, the unit cell period is $p=1.2l$ in both directions.
The dipoles are tilted by an angle $\alpha=20^\circ$ as compared with the ground plane, and their center is at the distance $d=0.25l$ above the ground plane.
Each element is fed in its center by a voltage-gap excitation, and the broadside radiation case is considered.
The comparison of different Q-factor methods is shown in Fig.~\ref{fig:Q_tilted_dipole}: our expression $\max (Q_{\rm e},Q_{\rm m})$ (dashed lines), the tuned bandwidth from CST simulation $Q_{\rm B}$ (magenta line), and the input-impedance-based Q-factor $Q_{\rm Z}$ calculated from impedance given by both our MoM code (black dotted curve) and CST simulation (green curve). 
Below the first grating lobe, located at $kl \simeq 5.2$, an overall good agreement between all the methods is observed; the proposed expression slightly overestimates the Q-factor at $kl \simeq 3 - 4.5$.
Above the electrical length $kl\simeq 5$, all methods give somewhat different Q-factor values, however, the second grating lobe at $kl \simeq 7.4$ is captured by all of the methods.
The differentces in the curves at the electric lengths between the two grating lobes are due to limited validity of the Q-factor description at low Q-values, as $Q_{\rm B}\simeq 5$ there.

\subsection{Beam scanning}
\begin{figure}[tb]
  \centering
  \begingroup
	\tikzset{every picture/.style={scale=1}}
	\includegraphics{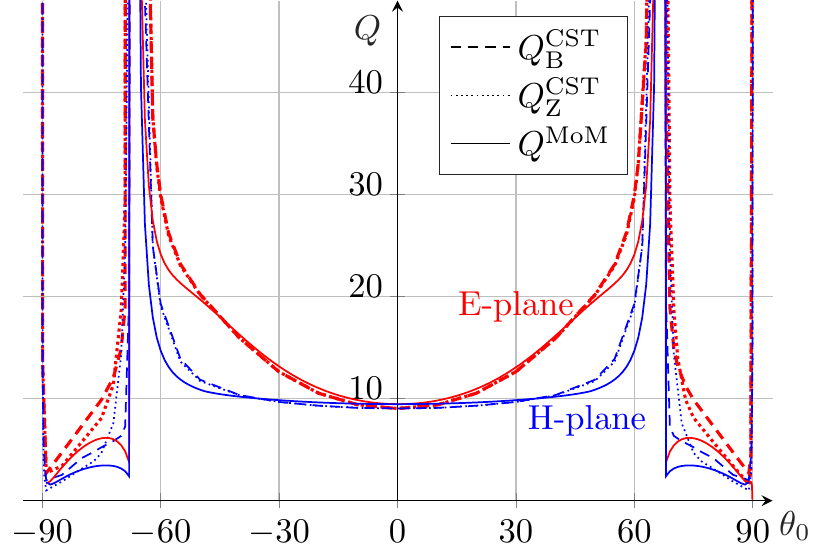}
  \endgroup
  ~
  \caption{The Q-factor of an array of tilted dipoles over a ground plane as a function of the scan angles at $kl\simeq 2.7$.}
  \label{fig:Q_steering}
\end{figure}

We consider the beam-scanning of the tilted dipoles array from the previous example at $kl\simeq 2.7$.
The beam-scanning angles, polar $\theta_0$ and azimuthal $\phi_0$, enter the Green's function~\eqref{eq:Greens_function} via the phase-shift vector $\Ktnull=k\sin\theta_0\cos\phi_0\hat{\vec{x}}+k\sin\theta_0\sin\phi_0\hat{\vec{y}}$.
The Q-factor as a function of the polar angle $\theta_0$ is shown in Fig.~\ref{fig:Q_steering} for E-plane ($\phi_0=0^\circ$, red curves) and H-plane ($\phi_0=90^\circ$, blue curves), where the azimuthal angle is counted from the positive $x$-semiaxis (see the inset of Fig.~\ref{fig:Q_tilted_dipole}).
The Q-factor $Q=\max(Q_{\rm e},Q_{\rm m})$ calculated by the proposed expression, is shown by solid curves, the tuned bandwidth, recalculated in the equivalent Q-factor $Q_{\rm B}$ is given by dashed lines, and the input-impedance Q-factor $Q_{\rm Z}$ is shown by dotted lines.
All the Q-factor methods agree reasonably well, the slight discrepancies between the proposed expression and the other two methods are observed around $\theta_0 = \pm 60^\circ$ in both planes.
The phased array blind spots at $\theta_0 = \pm 67^\circ$ are depicted by all the methods.
Due to the symmetry of the unit cell with respect to the $xz$-plane, the H-plane Q-factor is symmetric in $\theta_0$.
Although the unit cell is not symmetric with respect to the $yz$-plane, the asymmetry of the Q-factor in the E-plane is negligible.

\subsection{Square loop array}
\begin{figure}[tb]
  \centering
  \begingroup
	\tikzset{every picture/.style={scale=1}}
	\includegraphics{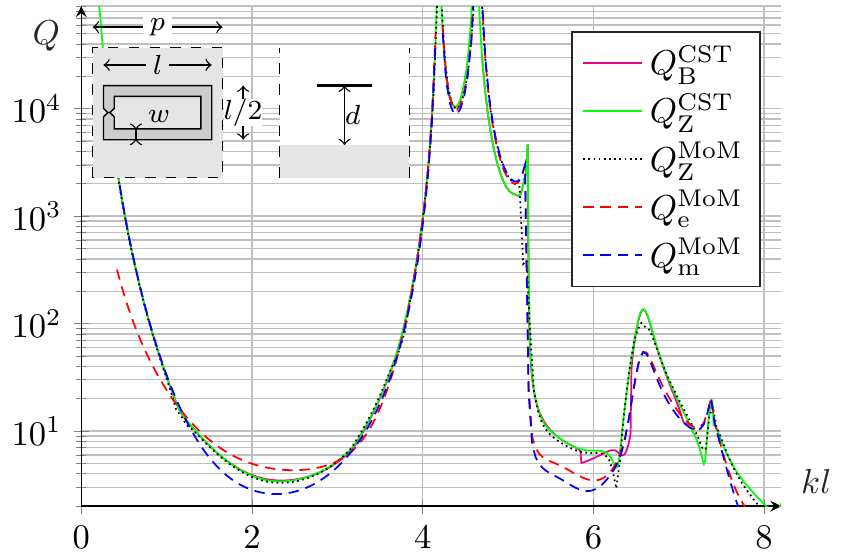}
  \endgroup
  ~
  \vspace{-12pt}
  \caption{The Q-factor of an array of rectangular loops over a ground plane. Strip width $l/10$, quadratic unit cell with $p=1.2l$, distance to ground plane $d=0.75l$.}
  \label{fig:Q_sloop}
\end{figure}

The Q-factor for an array of rectangular loops is shown in Fig.~\ref{fig:Q_sloop}.
The loop size is $l/2\times l$, strip width is $w=l/10$, unit cell period is $p=1.2l$, and the distance between the elements and the ground plane is $d=0.75l$, which corresponds to a quarter of the loops perimeter.
The loops are fed by a voltage gap excitation in a middle of one of its shorter sides, and the radiation is in the broadside direction.
The colors of the curves, corresponding to the different Q-factor methods are the same as in Fig.~\ref{fig:Q_tilted_dipole}.
All the methods agree below the grating lobe at $kl\simeq 5.2$ with a slight overestimate by the proposed method at $kl\simeq 1.5-3$. The curves are reasonably close above the first grating lobe.
The distance $d$ corresponds to a quarter-wavelength at $kl\simeq 2.1$, where the local minimum of the Q-factor is situated.
At $kl=4.2$, where $d$ is equal to a half wavelength, the presence of the ground plane results in a singular behavior of the Q-factor (compare with no ground plane case for the same element in~\cite{Osipov+Jonsson2019}).
The peak at $kl\simeq 4.6$ is related to the resonance of the loop~\cite{Osipov+Jonsson2019}.

\section{Conclusions}
\label{sec:conclusions}
In this paper, we derived the Q-factor expression for two-dimensionally periodic arrays over the ground plane.
It takes into account the exact shape of the array element, multiple propagating Floquet modes, and beam scanning.
The proposed expression is a representation in terms of current density on the array element only, while the ground plane currents are accounted for through the mirror method, and they are thus represented by additional terms in integral kernels of stored energy and radiated power.
Consequently, in the MoM numerical implementation only the array element needs to be discretized.
The here derived expression is an extension of our earlier result~\cite{Osipov+Jonsson2019}, where the arrays in free space were considered. 

The considered numerical examples indicate, that the proposed expression agrees well with the full-wave simulated and tuned bandwidth and with Yaghjian and Best input-impedance formula~\cite{Yaghjian+Best2005} below the grating lobes.
Above the grating lobes, where several propagating Floquet modes are present, the proposed Q-factor demonstrates a reasonable agreement with the other methods.
At this region, the performance of the different Q-factor methods is related to the validity of the Q-factor description.

\appendix

\section{Green's functions}

The two-dimensionally periodic free-space Green's function in the spectral form is~\cite{Oroskar+etal2006} 
\begin{equation}
\begin{split}
\G&(\vec{r}_1,\vec{r}_2) \\
&=\frac{1}{2\iu ab} 
\underset{(m,n)\in \Z^2}{\sum}
\frac{1}{\Kzmn}\eu^{-\iu\Ktmn\cdot(\vec{\rho}_1-\vec{\rho}_2)}\eu^{-\iu \Kzmn|z_1-z_2|},
\label{eq:Greens_function}
\end{split}
\end{equation}
with $\Ktmn=\Ktnull+2\pi\frac{n}{a}\hat{\vec{x}}+2\pi\frac{m}{b}\hat{\vec{y}}$, $\Kzmn=\sqrt[]{k^2-\Ktmn\cdot \Ktmn}$, and $z_i=\vr_i\cdot\vhat{z}$, $\vec{\rho}_i= \vr_i - \hat{\vec{z}}z_i$, $i=\{1,2\}$.
The part of the Green's function that is associated with the propagating modes is
\begin{equation}
\begin{split}
\Gp&(\vec{r}_1,\vec{r}_2) \\
=&\frac{1}{2\iu ab} \underset{(m,n)\in \mathcal{P}}{\sum}
\frac{1}{\Kzmn}\eu^{-\iu\Ktmn\cdot(\vec{\rho}_1-\vec{\rho}_2)}\eu^{-\iu \Kzmn|z_1-z_2|}.
\label{eq:Greens_function_F}
\end{split}
\end{equation}
where $\mathcal{P}=\{ (m,n): k^2-\Ktmn\cdot \Ktmn \geq 0 \}$ is the set of propagating modes.

The integral~\eqref{eq:g_kernel} reduces to~\cite{Osipov+Jonsson2019}  
\begin{equation}
\begin{split}
g(\vec{r}_1,\vec{r}_2) = 
&\frac{1}{4 ab} 
\underset{(m,n)\in \Z^2\setminus \mathcal{P}}{\sum}
\frac{1}{|\Kzmn|^2} 
\eu^{\iu \Ktmn\cdot(\vec{\rho}_1-\vec{\rho}_2)} \\
&\eu^{-|\Kzmn||z_2-z_1|} \left( \frac{1}{|\Kzmn|} + |z_1-z_2|  \right).
\label{eq:little_g_exact}
\end{split}
\end{equation}

%\bibliographystyle{IEEEtran}
%\bibliography{stored}

\end{document}